\documentclass[aps,prl,twocolumn,groupedaddress]{revtex4}
\usepackage{graphicx}

\begin{document}

\title{Resistively-Detected NMR in a Two-Dimensional Electron System near $\nu = 1$: Clues to the Origin of the Dispersive Lineshape}

\author{L. A. Tracy$^1$, J.~P. Eisenstein$^1$, L.~N. Pfeiffer$^2$, and K.~W. West$^2$}

\affiliation{$^1$California Institute of Technology, Pasadena CA 91125 
\\
$^2$Bell Laboratories, Lucent Technologies, Murray Hill, NJ 07974\\}

\date{\today}

\begin{abstract}
Resistively-detected NMR measurements on 2D electron systems near the $\nu = 1$ quantum Hall state are reported. In contrast to recent results of Gervais \emph{ et al.} [Phys. Rev. Lett. $\bf 94$, 196803 (2005)], a dispersive lineshape is found at all RF powers studied and Korringa-like nuclear spin-lattice relaxation is observed.  The shape of the unexplained dispersive lineshape is found to invert when the temperature derivative of the longitudinal resistance changes sign.  This suggests that both Zeeman and thermal effects are important to resistively-detected NMR in this regime.
\end{abstract}

\pacs{73.40.-c, 73.20.-r, 73.63.Hs}

\maketitle
Two dimensional electron systems (2DES) in semiconductors are weakly coupled to the nuclear magnetic moments of the host material via the hyperfine interaction.  This coupling allows for studies of the spin degree of freedom in the electronic system via nuclear magnetic resonance (NMR) techniques.  Among the several important findings that have stemmed from this connection, the discovery of multi-spin ``skyrmion'' excitations in the quantum Hall effect (QHE) regime is particularly significant\cite{barrett1}.  Resistively-detected NMR (RDNMR), in which the resistance of the 2DES is modified by NMR excitation, has also led to intriguing observations, including signatures of competition between collective electronic phases with different electronic spin configurations\cite{kronmuller1,smet1,stern,spielman1}. 

Typically, the hyperfine coupling between nuclear moments and 2D electron spins is expressed in terms of an effective magnetic field $B_N$.  This field, which does not influence the orbital motion of the 2D electrons, contributes to their spin Zeeman energy: $E_Z = g\mu_B(B+B_N)$, with $g$ the electron $g$-factor, $\mu_B$ the Bohr magneton, and $B$ the externally applied magnetic field.  $B_N$ depends upon the nuclear spin polarization and can reach several tesla in GaAs.  Consequently, it is often assumed that, at least for electronic phases in which the Zeeman energy contributes to the energetics and transport in the 2DES, modification of $B_N$ via radio-frequency (RF) excitation is what enables RDNMR.  Interestingly, $B_N$ and $B$ have opposite signs in GaAs, owing to the negative $g$-factor ($g \sim -0.4$) of electrons in the conduction band. As a result, destruction of nuclear polarization via NMR excitation usually $increases$ the electronic Zeeman energy.

NMR-induced modification of $E_Z$ in 2D systems is almost certainly the origin of some of the RDNMR responses which have been reported\cite{kronmuller1,smet1,stern,spielman1}.  However, more complex RDNMR effects have also been found.  Desrat, {\it et al.}\cite{desrat} reported a curious ``dispersive'' RDNMR lineshape (vs. frequency) near the integer QHE state with one fully occupied spin-resolved Landau level (i.e. near filling factor $\nu = 1$).  They found that in slow frequency sweeps through the NMR line the longitudinal resistance $R_{xx}$ displays roughly equal positive and negative excursions from its equilibrium value.  Desrat, {\it et al.} suggested that this unusual resonance shape might be due to the formation of a skyrmion lattice.  Interestingly, dispersive lineshapes have also been reported at certain other filling factors\cite{stern,gervais2}.

Here we report RDNMR studies of 2D electron systems in the QHE regime, focussing on the regions surrounding $\nu = 1$. In agreement with Desrat, {\it et al.}\cite{desrat}, we find that the equilibrium RDNMR response has a dispersive frequency dependence near $\nu = 1$.  While the amplitude of this response evolves continuously with RF power, its dispersive shape remains the same down to the lowest powers studied.  Using a frequency-jumping technique we determine the nuclear spin-lattice relaxation rate $T_1^{-1}$ near $\nu = 1$.  We find that $T_1^{-1}$ exhibits a Korringa-like temperature dependence near $\nu = 1$. Most interestingly, we find that the shape of the dispersive RDNMR line inverts at a well-defined filling factor on the flank of the $\nu = 1$ QHE.  Remarkably, this shape inversion coincides with a change in sign of $dR_{xx}/dT$, the temperature derivative of the longitudinal resistance.
This suggests that Zeeman effects are not solely responsible for RDNMR near $\nu = 1$ and that thermal effects also play a role. 

The samples used in the present experiments are modulation-doped GaAs/AlGaAs heterostructures containing high mobility 2DESs.  For the data presented here, the 2DES is confined in GaAs at a single interface with AlGaAs and is laterally patterned into a wide (500 $\mu$m) Hall bar geometry.  The density of the 2DES is $N_s \approx 1.6 \times 10^{11} \rm cm^{-2}$ and its low temperature mobility is $\mu \approx 8 \times 10^6 \rm cm^2/Vs$.  Diffused In ohmic contacts enable low frequency (13 Hz, typically) magneto-transport measurements.  Cooling of the 2DES occurs primarily through the well-heat-sunk (and filtered) Au wires attached to these contacts.  Importantly, the thermal relaxation time of the 2DES (measured via pulsed ohmic heating experiments) is quite short, less than 0.1 sec. throughout the regime of these experiments.

An approximately rectangular 8-turn NMR coil is wound around the sample for applying a RF magnetic field $H_1$ parallel to the 2DES plane and perpendicular to the applied dc magnetic field.  We estimate $H_1$ to be in the 0.1 –- 0.5 $\mu$T range\cite{H1estimate}.  These RF fields are far smaller than the typical nuclear dipolar field, $H_d \sim 100 ~\mu$T. Heating induced by the RF was readily calibrated out using the measured temperature dependence of the 2DES resistivity; all temperatures quoted here have been corrected for this effect.

\begin{figure}[t]
\centering
\includegraphics[width=3.5 in,bb= 71 72 341 260]{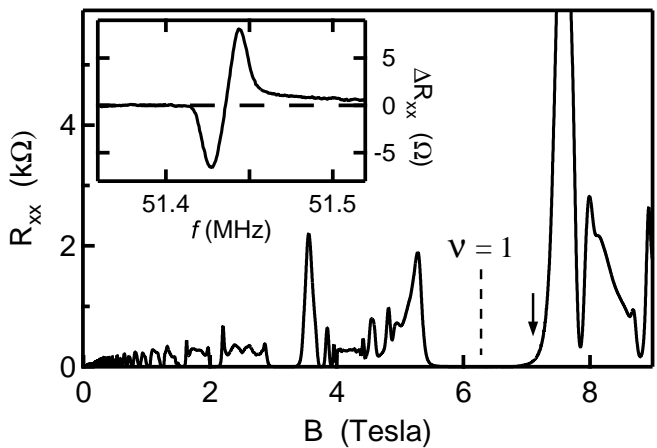}
\caption{Longitudinal resistance $R_{xx}$ vs. magnetic field. Dashed line indicates center of $\nu = 1$ QHE.  Inset: Resistively-detected $^{75}$As NMR at $B = 7.1$ T, denoted by arrow in main panel. All data at $T$ = 70 mK.}
\label{fig1}
\end{figure}
Figure 1 displays the longitudinal resistance $R_{xx}$ in the present sample as a function of perpendicular magnetic field at $T = 70$ mK.  The broad minimum centered around $B = 6.4$ T reflects the $\nu = 1$ QHE state in which the 2DES density $N_s$ matches the degeneracy $eB/h$ of the lowest spin-resolved Landau level.  The inset to Fig. 1 shows the response of $R_{xx}$ to a slow sweep of the frequency of RF excitation of the NMR coil surrounding the sample.  For these data the magnetic field is set to $B = 7.1$ T, i.e. on the high field flank of the $\nu = 1$ QHE minimum.  At this field $\nu \approx 0.90$ and $R_{xx}$ is just becoming significant as quasi-holes in the lowest Landau level delocalize and begin to conduct.  A strong resonant response is apparent in the figure and corresponds to NMR of the $^{75}$As nuclei in the sample.  As first observed by Desrat, {\it et al.}\cite{desrat} the resonance has a dispersive, or ``derivative'' shape: $R_{xx}$ falls below its equilibrium value on the low frequency side of the resonance and then rises above it on the high frequency side. The approximately 15 kHz width of the resonance is about one order of magnitude larger than that expected from simple nuclear dipolar broadening. The distribution of Knight shifts resulting from the shape of the 2D electron subband wavefunction is almost certainly a significant contributor to the observed width\cite{barrett1}. Varying the excitation current $I$ used for the magneto-transport measurement from 2 nA to 100 nA produced no qualitative effect on the shape of these resonances.  We emphasize that while dispersive lineshapes like that in Fig. 1 are also found on the low field flank of the $\nu = 1$ QHE, they are by no means ubiquitous in the QHE regime. Non-dispersive, or ``conventional'' RDNMR lineshapes are more commonly seen\cite{smet1,stern,desrat,hashimoto}.  For example, at $\nu = 1/2$ we and others have observed simple, unipolar RDNMR lineshapes\cite{stern,spielman1,tracy}.

The RDNMR spectrum shown in Fig. 1 was obtained by sweeping the RF frequency slowly (0.13 kHz/sec) through the line. At this rate the difference between sweeping the frequency up vs. down is small. RDNMR spectra can also be obtained by abruptly adjusting the frequency to a value $f$ within the resonance region from a value $f_0$ several linewidths away\cite{desrat}.  The resistance $R_{xx}$ responds by relaxing to a new value. After equilibrium is reached, the frequency is then reset to $f_0$ and $R_{xx}$ returns to its off-resonance value.  Collecting a series of such transients, with different on-resonance frequencies $f$, allows construction of the entire RDNMR spectrum.  Good agreement with slowly swept spectra like that in Fig. 1 is obtained.  This frequency-jumping technique also allows for examination of the dependence of the RDNMR signal $\Delta R_{xx}/R_{xx}$ on the RF power delivered to the NMR coil. Figure 2 shows this power dependence in a typical dispersive RDNMR line at the two frequencies where $\Delta R_{xx}/R_{xx}$ is maximally positive and negative.  
At the maximum power shown we estimate $H_1 \approx 0.5~\mu$T.  In addition to demonstrating that the nuclear spin system is not saturated, these data prove that the dispersive character of the RDNMR lineshape in our sample persists down to the lowest powers used ($H_1 \approx 0.1~\mu$T).

\begin{figure}[t]
\centering
\includegraphics[width=3.3 in,bb= 68 72 341 245]{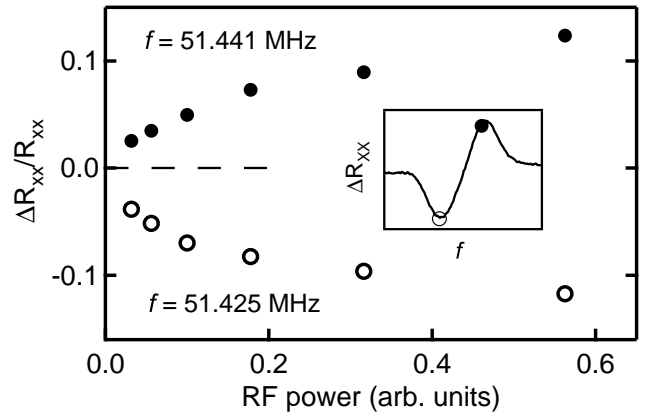}
\caption{RF power dependence of RDNMR signals at $\nu = 0.88$ and $T \approx 65$ mK.  Inset depicts frequencies on dispersive RDNMR lineshape where data was acquired.}
\label{fig2}
\end{figure}

The transient relaxation of $R_{xx}$ following the return of the RF frequency $f$ to its off-resonance value $f_0$ offers a simple way to assess the nuclear spin lattice relaxation rate $T_1^{-1}$.  We find that $T_1^{-1}$ varies monotonically with frequency across the RDNMR line, faster rates being observed on the low frequency side of the line.  This variation is not surprising given the distribution of Knight shifts arising from the shape of the 2D electron subband wavefunction. The low frequency side of the RDNMR resonance corresponds to the largest Knight shifts and thus derives from those nuclei lying near the maximum of the subband wavefunction. These same nuclei will relax faster than those in locations where the subband wavefunction is small, consistent with our observations.  While the transient responses of $R_{xx}$ when jumping off the low frequency side of the RDNMR line are well-fit by a simple exponential, this is not the case on the high frequency side.  Here the transients exhibit a long tail which we speculate arises from nuclei outside the quantum well whose influence is felt via slow nuclear spin diffusion. In the following we report $T_1$ values obtained from the low frequency side of the line, at the frequency where the RDNMR response is maximally negative\cite{T1}.

In agreement with previous work\cite{barrettT1,smet2,hashimoto}, we find short $T_1$ times ($\sim$5 sec. at $T = 50$ mK being typical) near $\nu = 1$, with local maxima in the rate $T_1^{-1}$ both below and above $\nu = 1$. $\rm C\hat{o} t \acute{e}$, {\it et al.}\cite{cote} have attributed this fast relaxation to low-lying spin-wave modes of a skyrmion solid and used it to explain the enhanced heat capacity reported by Bayot, {\it et al.}\cite{bayot}.  The maximum in $T_1^{-1}$ on the low $\nu$ side of $\nu = 1$ at $T = 70$ mK is shown in Fig. 3a. At each point in Fig. 3a a dispersive RDNMR lineshape is observed.  Figure 3b shows the temperature dependence of $T_1^{-1}$ at $\nu = 0.88$, i.e. near the maximum in $T_1^{-1}$ vs. $\nu$. The data reveal that $T_1^{-1}$ is linear in temperature from $T = 120$ mK down to about 45 mK.  The solid line in the figure corresponds to the best-fit Korringa law: $T_1T$ = 0.28 sec-K.  This temperature dependence is consistent with the theory of $\rm C\hat{o} t \acute{e}$, {\it et al.}\cite{cote}.  

Recently, Gervais, {\it et al.}\cite{horst} have reported RDNMR measurements near $\nu = 1$ on a 2DES with mobility $17 \times 10^6 \rm ~cm^2/Vs$ and density $1.6 \times 10^{11} ~\rm cm^{-2}$, confined to a 40 nm GaAs quantum well. They find a dispersive lineshape only at high RF powers. For RF powers corresponding to $H_1 \approx 1 ~\mu$T they instead find that $\Delta R_{xx}$ shows only negative excursions as the RF frequency is swept through resonance.  In contrast, we find a dispersive lineshape down to $H_1 \approx 0.1 ~\mu$T. Gervais, {\it et al.} also report long $T_1$ times (from $\sim 20$ to 600 sec.) and that $T_1^{-1}$ \emph{rises} with falling temperature, in the same regime of temperature and filling factor where we find the opposite behavior.  Although these discrepancies might be related to sample and/or measurement differences, we have reproduced our basic findings on the lineshape and the $T_1$ temperature dependence using a 2DES with mobility $14 \times 10^6 \rm ~cm^2/Vs$, confined to a 30 nm GaAs quantum well.

\begin{figure}[t]
\centering
\includegraphics[width=3.5 in,bb= 74 65 386 242]{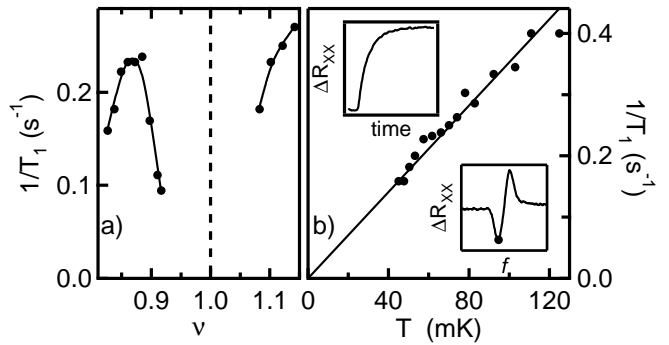}
\caption{a) $T_1^{-1}$ vs. filling factor near $\nu = 1$ at $T = 70$ mK. Lines are guides to the eye. b) $T_1^{-1}$ vs. temperature at $\nu = 0.88$. Solid line: Korringa law fit; $T_1T = 0.28$ sec-K. Upper inset: Typical RDNMR transient following frequency jump: $f \rightarrow f_0$. Lower inset: Dot shows position on dispersive lineshape where $T_1$ was measured.}
\label{fig3}
\end{figure}
A dispersive lineshape is inconsistent with the usual model in which RDNMR is due solely to a hyperfine-induced increase of the electronic Zeeman energy.  The low temperature transport properties of the $\nu = 1$ QHE are believed to reflect the existence of skyrmionic quasiparticles, the energy gap for which increases monotonically with $E_Z$.  As a result, the $\nu = 1$ QHE should be strengthened by an NMR-induced increase in $E_Z$ and the resistance $R_{xx}$ exhibit a simple minimum vs. frequency.  

Alternatively, the dispersive RDNMR lineshape might reflect the temperature of the electron gas, driven out of equilibrium via interaction with the nuclear spin system and the RF electromagnetic field.  Since $dR_{xx}/dT > 0$ near well-developed QHE states, the data in Fig. 1 could imply that the electron gas is colder than the background thermal reservoir (crystal lattice, metallic ohmic contacts, etc.) on the low frequency side of the NMR line, but hotter than it on the high frequency side.  

To investigate this possibility, we have examined the relationship between the dispersive RDNMR resonance and the magnitude and sign of $dR_{xx}/dT$.  While $dR_{xx}/dT > 0$ in the immediate magnetic field vicinity of $\nu = 1$, it crosses zero and changes sign on moving further away.  Sign changes such as these are commonplace in the QHE regime, although a complete understanding of them does not exist.  If the RDNMR lineshape reflects the electron temperature, its shape ought to invert when $dR_{xx}/dT$ changes sign.  Figure 4 demonstrates that this is precisely what occurs.  In Fig. 4a the $R_{xx}$ peak on the high magnetic field flank of the $\nu = 1$ QHE is shown at three closely-spaced temperatures: 62, 65, and 70 mK.  A clear sign change in $dR_{xx}/dT$ is observed at $B \approx 7.65$ T.  Figs. 4b and 4c display RDNMR spectra at $B =$ 7.6 and 7.7 T, fields which straddle the sign change in $dR_{xx}/dT$.  The two spectra have clearly inverted shapes.  In both cases, the electron gas appears to cool on the low frequency side of the resonance and to warm on the high frequency side.  Not surprisingly, RDNMR signals are very weak closer to the sign change where $dR_{xx}/dT \approx 0$. 

\begin{figure}[t]
\centering
\includegraphics[width=3.5 in,bb= 86 209 430 441]{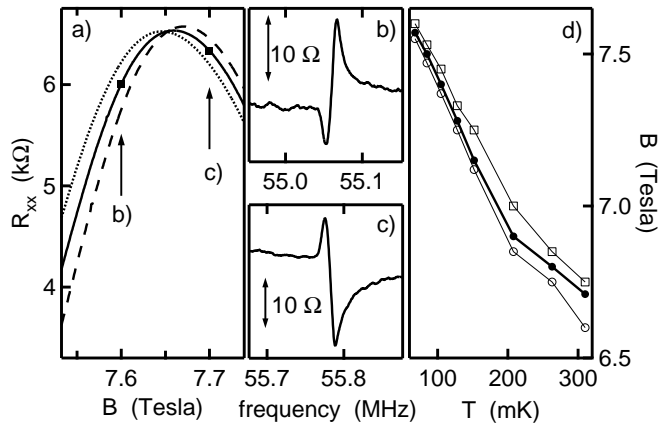}
\caption{a) $R_{xx}$ peak on high field flank of $\nu = 1$ QHE at $T =$ 62 mK (dashed), 65 mK (solid), and 70 mK (dotted). Note change in sign of $dR_{xx}/dT$ near $B = 7.65$ T.  Arrows indicate the magnetic fields, $B =$ 7.6 and 7.7 T, at which the RDNMR resonances shown in b) and c) were obtained. d) Solid dots: Magnetic field where $dR_{xx}/dT$ changes sign, versus temperature\cite{cooldown}.  Open circles and squares are fields where RDNMR spectra with inverted shapes are observed.}
\label{fig4}
\end{figure}
At higher temperatures the QHE minimum in $R_{xx}$ narrows and the field where $dR_{xx}/dT$ changes sign shifts.  Figure 4d shows that the RDNMR shape inversion faithfully tracks these shifts\cite{cooldown}; the solid dots give the magnetic field of the sign change while the open circles and squares are fields where RDNMR spectra with the same basic inverted shapes shown in Fig. 4b and 4c, respectively, are observed.  These fields closely straddle the location of the sign change and thus strongly suggest that the RDNMR shape inversion and the $dR_{xx}/dT$ sign change are coincident.  We emphasize that while the $T \approx 65$ mK data in Fig. 4a suggest that the $dR_{xx}/dT$ sign change coincides with the maximum in $R_{xx}$, this is not true in general.  At higher temperatures the sign change occurs at a significantly lower field than the maximum in $R_{xx}$.  From these several observations we conclude that the RDNMR spectra are essentially proportional to $dR_{xx}/dT$, at least in the vicinity of $\nu =1$.  This supports the idea that the RDNMR lineshape reflects small ($\sim 1$ mK) temperature changes of the 2DES. 

A model which might account for these observations is built around the idea that RF photons, detuned slightly from resonance with the nuclear Larmor frequency, can nonetheless induce nuclear spin flips if the energy mismatch is accomodated by the 2D electron gas\cite{girvin}.  For RF frequencies $\omega_{RF}$ less than the nuclear Larmor frequency $\omega_N$ (positive detuning) the electron gas must supply energy to the nuclear system.  As a result, cooling of the electrons occurs.  For negative detuning, $\omega_{RF} > \omega_N$, the electron gas absorbs the excess energy and heats up. However, it is difficult to reconcile this model with the usual descriptions of non-saturated NMR under low power continuous RF illumination. Since the dispersive RDNMR lineshape is independent of the frequency sweep rate (so long as it is sufficiently slow), a steady-state description ought to be possible.  In this case the Bloch equations reveal that the nuclear spins are heated out of equilibrium and energy is flowing into the lattice degrees of freedom, including the 2D electron system.  Cooling of the electron gas thus seems unlikely.  

An alternative model of the dispersive lineshape recognizes that NMR-induced heating of the nuclear spin system has two distinct effects on the electron gas.  One is the increase in the electronic Zeeman energy described in the introduction.  In addition, however, the hot nuclei will naturally heat the electron gas to some degree.   Near the $\nu = 1$ QHE the former effect reduces $R_{xx}$ while the latter increases it.  Importantly, the two effects have different dependences on RF frequency.  The Zeeman effect, just like the Knight shift, is largest on the low frequency side of the NMR line. In contrast, the NMR-induced heating of the electron gas is roughly independent of frequency across the NMR line. These different frequency dependences could thus explain the dispersive shape of the RDNMR line.  But this model also has flaws: While the heating effect on $R_{xx}$ will change sign with $dR_{xx}/dT$ it not obvious why the Zeeman effect would.  We can only comment that the Zeeman energy dependence of $R_{xx}$ is well understood only close to $\nu = 1$; how it behaves at larger $|\nu –- 1|$ is unknown.   

In conclusion, we have examined resistively-detected NMR near the $\nu = 1$ quantized Hall effect.  We find a dispersive lineshape at all RF power levels.  Accurate measurements of the nuclear spin-lattice relaxation time are in good agreement with theoretical expectations.  Surprisingly, we find the shape of the RDNMR line inverts when the temperature derivative of the longitudinal resistance, $dR_{xx}/dT$, changes sign.  While this is not yet understood, it strongly suggests that both Zeeman and thermal effects are important in RDNMR near $\nu = 1$.  

It is a pleasure to acknowledge several helpful conversations with S.M. Girvin.  This work was supported by the DOE under Grant No. DE-FG03-99ER45766 and the NSF under Grant No. DMR-0242946.


\begin{references}
\bibitem{H1estimate} These estimates were derived from a careful circuit analysis of the 50-ohm coax + NMR coil system.
\bibitem{cooldown} The data in Fig. 4d was taken after thermally cycling the sample to room temperature.  This produced a small density shift ($\sim 1\%$) and similarly small shifts in the field location where $dR_{xx}/dT$ changes sign.
\bibitem{T1} The observed RDNMR transient response times exceed the electronic thermal relaxation time by at least a factor of 25.  This justifies our association of the transient response time with the nuclear $T_1$.
\end{references}
\end{document}